\documentclass[a4paper,11pt]{article}
\usepackage{latexsym}

\begin{document}

\author{M.Mekhfi \thanks{%
Mailing address:Laboratoire des sciences de la mati\`{e}re condens\'{e}e, D%
\'{e}partement de physique {}Universit\'{e} Es-senia 31100 Oran ALGERIE} 
\thanks{{\footnotesize mekhfi@myrealbox.com}}}
\title{Cohomology and Bessel Functions Theory}
\maketitle

\begin{abstract}
By studying cohomological quantum mechanics on the punctured plane,we were
led to identify (reduced ) Bessel functions with homotopic loops living on
the plane.This identification led us to correspondence rules between
exponentials and Bessel functions.The use of these rules makes us retrieve
known but also new formulas in Bessel functions theory.
\end{abstract}

\section{Cohomology and topological actions}

A recall of few basic notions of cohomology together with associated
topological actions and topological quantum mechanical systems is necessary
in order to understand the relationships of these backgrounds to the theory
of Bessel functions.Let $M$ be a compact manifold with local coordinates $%
x^{\mu }$ and let $H^{r}(M)$ be the $r^{th}$ de Rham cohomology group.Let $%
c_{1,}.........c_{k}$ be elements of the homology group $H_{r}(M)$ with $k$
the $r^{th}$ Betti number and of the same class $[c_{i}]=[c_{j}].$Then for
any set of numbers $b_{1},......b_{k}$ a corollary of the de Rham's theorem
states that there exist a closed $r$-form $\omega $ such that

\begin{equation}
\int_{c_{i}}\omega =b_{i} \textrm{ \qquad }1\leq i\leq k  \label{1}
\end{equation}

We may include in this expression the trivial period $b_{i}=0$ which
corresponds to $\omega $ being a closed and exact form.The numbers in \ref{1}
are the periods of closed $r$-forms over cycles $c_{i}.$The main features of
such actions is that there are defined on the product H$_{r}$(M)$\times $H$%
^{r}$(M) and are therefore topological actions,that is invariant under any
infinitesimal deformations which keeps the cycle within its homologyclass $%
H^{r}(M)$

\[
\delta x^{\mu }=\epsilon ^{\mu } 
\]

Topological (cohomological) quantum mechanical systems we are interested in
are those described by the topological actions \ref{1} .The application to
Bessel functions comes from considerations of the first non trivial action
with period $b_{1}$on the punctured plane .The topological ingredients are
loops and 1-forms .

A very practical way to gauge fix these actions is to use BRST symmetry.To
the deformation symmetry at hand we associate the BRST symmetry with
generator $s$

\begin{eqnarray*}
sx^{\mu } &=&\psi ^{\mu } \\
s\psi ^{\mu } &=&0 \\
s\bar{\psi}^{\mu } &=&\lambda ^{\mu } \\
s\lambda ^{\mu } &=&0
\end{eqnarray*}

To gauge fix the symmetry such that covariance is maintained and to get an
action quadratic in velocities Baulieu and Singer proposed the following
gauge function $\dot{x}^{\mu }+\frac{\partial V}{\partial x^{\mu }}+\cdots $
where $\dot{x}^{\mu }=\frac{dx^{\mu }}{dt}$ and where we omit the
Christoffel symbol term as these are known matters and not relevant to what
follows.The important ingredient here is the prepotential $V$ .It is a
priori an arbitrary function of the coordinates $x^{\mu }$ but to correctly
define topological invariants one has to define the prepotential
properly.First write the gauge fixed action

\begin{equation}
S_{GF}=\int_{c\in H_{1}}\omega +\int dt\!\quad \!\!s\bar{\psi}^{\mu }(g_{\mu
\nu }\dot{x}^{\nu }+\frac{\partial V}{\partial x^{\mu }}-\frac{1}{2}g_{\mu
\nu }\lambda ^{\mu }+\cdots  \label{4}
\end{equation}

After integrating out the auxiliary field $\lambda $ we select out the
bosonic linear term of interest.

\[
\int dt\,\dot{x}^{\mu }\frac{\partial V}{\partial x^{\mu }}=\int dV 
\]

The second step is to cancel out the classical action.This cancellation is
necessary for the resulting gauge fixed action to possess a secondary (dual)
BRST symmetry $\bar{s}$ which in turn allows us to write the hamiltonian in
the form $H\propto \{Q,\ \bar{Q}\}$ with $Q$ and $\ \bar{Q}$ both
nilpotent.The proper definition of the prepotential is therefore given by
the equation

\begin{equation}
\int_{c\in H_{1}}\omega =\int_{c\in H_{1}}dV  \label{6}
\end{equation}

In other word one should not simply look for a prepotential on the basis
that the integral of it numerically cancels the classical action but shoud
first define a period (here $b_{1}$).We hereafter will restrict to the
punctured plane $R^{2}/(0)$and take it as our target manifold \ as it is
simple with a non trivial homology and specially because it has a direct
application to the theory of Bessel functions on concern in this paper.We
thus have $H^{1}(R^{2}/(0),R)\cong R$ that is $\omega ^{`s}$are one forms
labelled by real numbers and the cycles $c_{i}$ are homotopic loops
encircling the whole as $H_{1}\cong \Pi \cong Z$ where $\Pi $ is the
homotopy group or the fundamental group and $Z$ is the set of integers.The
topological action associated with the punctured plane is ( $\lambda \in R$
and $x^{1}+ix^{2}=r\exp (i\theta )$)

\[
S_{cl}=\lambda \int_{c}d\theta 
\]

The complete solution to the equation \ref{6} for the punctured plane is

\begin{equation}
V=\lambda (\theta +\Phi (\theta ))  \label{8}
\end{equation}

Where the function $\Phi (\theta )$ is any function but periodic $.$The
simplest case without $\Phi (\theta )\ $has been selected by Baulieu and
Rabinovici \cite{Chamseddine}.Such truncated solution neither lead to a
complete description of the invariants on the punctured plane nor to the
theory of Bessel functions .Let us remember that the topological invariants
on the punctured plane are $W$ and $\Pi (m)$ $m\in Z$ which are respectively
the winding number operator and the fundamental group operator ( $m$
indexing the group elements).They are defined as follows\footnote{%
We may use the representation$\mid \theta \ \rangle $ instead of $\mid n\
\rangle .$In this case we have the actions $\ $%
\begin{eqnarray*}
W &\mid &\theta \rangle =-i\partial _{\theta }\mid \theta \ \rangle \\
\Pi (m) &\mid &\theta \ \rangle =\exp im\theta \mid \theta \ \rangle
\end{eqnarray*}%
}

\begin{eqnarray*}
W &\mid &n\rangle =n\mid n\rangle \\
\Pi (m) &\mid &n\rangle =\mid n+m\rangle
\end{eqnarray*}

Where the state $\mid n\rangle $ describes the \textquotedblright
quantum\textquotedblright\ state of the loop encircling the whole $n$
times.The topological invariant which came out of the analysis of reference 
\cite{Chamseddine} is

\[
W+\lambda \Pi (0) 
\]

where $\Pi (0)=1$ is the group identity .The other group elements $\Pi (m)$ $%
m\neq 0$ are missing.We have completed their analysis \cite{Wissale} $\cite%
{Mus}$ by adding the missing part $\Phi (\theta )$ in the complete solution %
\ref{8} . The topological invariant turned out to be in this case

\[
W+\lambda \sum_{m\in Z}\Pi (m) 
\]

Where all the invariants are represented.The presence of the periodic
function $\Phi $ in $V$ is primordial and is at the heart of the connection
to Bessel function.

\section{Bessel functions are realization of homotopic loops}

The hamiltonian associated with the general action \ref{4} and adapted to
the punctured plane is

\begin{equation}
H=\frac{1}{2}p^{2}+\frac{1}{2r^{2}}(\frac{\partial V}{\partial \theta })^{2}-%
\frac{1}{2}[\bar{\psi}_{i},\psi _{j}]\frac{\partial ^{2}V}{\partial
x_{i}\partial x_{j}}=\frac{1}{2}\{Q,\ \bar{Q}\}  \nonumber
\end{equation}

Where $P_{i}=-i\frac{\partial }{\partial x_{i}}$ and $\bar{\psi}_{i}=\frac{%
\partial }{\partial \psi _{i}}$ are canonical momenta for the coordinates $%
x_{i}$ and the ghost fields $\psi _{i}$ respectively and where the
generators $Q$ and $\bar{Q}$ are given by

\begin{eqnarray*}
Q &=&\psi _{i}(p_{i}+i\frac{\partial V}{\partial x_{i}}) \\
\bar{Q} &=&\bar{\psi}_{i}(p_{i}-i\frac{\partial V}{\partial x_{i}})
\end{eqnarray*}

Let us note that these generators are nilpotent ,that is $Q^{2}=\bar{Q}%
^{2}=0 $ this is because there are built out of the anticommuting variables $%
\psi (\bar{\psi})$.In fact one can identify the first term in $Q$ ($p_{i}$
term )and $\bar{Q}$ respectively with the exterior derivative $d$ ( $%
d^{2}=0) $ and its adjoint $d^{\dagger }(d^{\dagger 2}=0)$ and the generator 
$Q$ itself with the twisted or deformed exterior derivative $d_{\alpha
}=\exp (\alpha V) $ $d$ $\exp (-\alpha V)$ first introduced by E.Witten \cite%
{Mohsine} and which served to investigate Morse theory in a very novel and
deep way.Let us rewrite $Q$ in the form of exponential as $d_{\alpha }$ by
using the relation $p_{i}+i\frac{\partial V}{\partial x_{i}}=\exp
(V)\,p_{i}\exp (-V)$ .Doing this we get

\begin{eqnarray*}
Q &=&\exp (V)\,Q_{0}\exp (-V) \\
\bar{Q} &=&\exp (-V)\,Q_{0}\exp (V)
\end{eqnarray*}

Where $Q_{0}$ is identified with the exterior derivative $d$ .

Bessel functions will enter in play when we consider the eigenvalue problem
associated to the Hamiltonian.The more elegant way to study the above
hamiltonian is to use the superwavefunction formalism \cite{Wissale}.Let $%
\Phi (x,\psi )$denotes the superwavefunction

\[
\Phi (x,\psi )=\phi +\psi _{i}A^{i}+\frac{1}{2}\in _{ij}\psi _{i}\psi _{j}B 
\]

Where the four states $\phi ,A^{i}$and $B$ are functions of the coordinates $%
x_{i}$ only.We will only retain ghost free solutions i.e. $\phi ^{`s}$
.These are eigenstates of the hamiltonian ( The ghost solutions will
probably give no direct informations on Bessel functioms !!)

\begin{eqnarray}
H\phi &=&\phi  \label{10.} \\
H &=&-\frac{\partial ^{2}}{\partial x_{i}\partial x^{i}}+\frac{1}{2r^{2}}[(%
\frac{\partial V}{\partial \theta })^{2}-\frac{\partial ^{2}V}{\partial
^{2}\theta }]  \nonumber
\end{eqnarray}

Let us look for a solution of the form

\[
{\large F}(r)\mathrm{f(\theta )} 
\]

Putting this into the $\phi $ equation \ref{10.} and separating the
variables,we get ($W=(\frac{\partial \Phi }{\partial \theta })^{2}-\frac{%
\partial ^{2}\Phi }{\partial ^{2}\theta }$)

\begin{eqnarray*}
0 &=&\mathrm{\ddot{f}}(\theta )+(\zeta ^{2}-\lambda ^{2}-W)\mathrm{f}(\theta
) \\
0 &=&{\large \ddot{F}}(r)+\frac{1}{r}{\large \dot{F}}+(-\frac{\zeta ^{2}}{%
r^{2}}+2E){\large F}
\end{eqnarray*}

Where $\zeta $ is the separation parameter.The $r$- component equation is
the differential equation defining Bessel functions $J_{\zeta }(\sqrt{2}Er)$%
,while the $\theta $-component equation is of Sturn-Liouville type.This
equation has been analyzed in \cite{Wissale} and put on the simplified form

\[
\ddot{u}+(\zeta ^{2}-\lambda ^{2}-W\,)\,u=0 
\]

Where the solutions are here periodic.There is a set of theorems of
eigenvalues and boundaries problems which state that the above equation
independently of the potential $W$ provided it is continuous,has real
eigenvalues and are all ordered as $0\langle \zeta _{1}^{2}\langle \cdots
\zeta _{p}^{2}$ and $\lim_{p\rightarrow \infty }\zeta _{p}=\infty $ and
moreover that the eigenfunctions $u_{\zeta _{p}}(\theta )$ are orthonormal
with weight 1,that is $\int u_{\zeta _{q}}^{\ast }(\theta )u_{\zeta
_{p}}(\theta )$ $d\theta =\delta _{qp}.$( These are generalization of $\exp
[i\sqrt{\zeta ^{2}-\lambda ^{2}}\theta ]$ ) when $W=0$ ).

We are now ready to write down expectation values for topological
invariants.On dimensional grounds ,one may select the following
candidates,together with their hermitian conjugates .\footnote{%
The topological invariant $-i\partial _{\theta }$ $+i\partial _{\theta
}V=-i\partial _{\theta }$ $+\lambda \sum_{m\in Z}\phi _{m}\exp (im\theta )$
can be written in an equivalent way in the $\mid n\rangle $ states as $%
W+\lambda \sum_{m\in Z}\Pi (m)$ for specific values of the $\phi _{m}^{`s}.$}

\begin{eqnarray*}
\{Q,\epsilon ^{ij}x_{i}\bar{\psi}_{j}\} &=&-i\partial _{\theta }+i\partial
_{\theta }V \\
\{Q,x_{i}\bar{\psi}_{j}\} &=&r\partial _{r}
\end{eqnarray*}

In writing these expressions we dropped out ghost terms as they give
vanishing actions on the $\phi $ eigenfunctions of zero ghost numbers we are
considering.The expectation value of the $\theta $-component topological
invariant is ( we only include the $\theta $-component of the wave function
since the operator only depends on $\theta $)

\begin{eqnarray*}
\langle \{Q,\epsilon ^{ij}x_{i}\bar{\psi}_{j}\}\rangle _{E,\zeta _{p}} &=&%
\frac{\int u_{\zeta p}^{\ast }(-i\partial _{\theta }+i\partial _{\theta
}V)u_{\zeta _{p}}}{\int u_{\zeta p}^{\ast }u_{\zeta _{p}}d\theta }=\langle
-i\partial _{\theta }+i\partial _{\theta }V\rangle ^{\zeta _{p}} \\
&=&\sqrt{\zeta _{p}^{2}-\lambda ^{2}}+\textrm{higher order in }V \\
&=&\zeta _{p}+\cdots
\end{eqnarray*}

This is an effective winding number which is $\zeta _{p}$ corrected by the
interaction $V$.

To define the second topological invariant one needs to interpret Bessel
functions as describing homotopic loops .To this end we have to restrict
ourselves to non interacting case i.e. $\lambda =0;W=0;\zeta _{p}=n$ .In
this case the wavefunctions are simple

\begin{eqnarray*}
u_{n}(\theta ) &=&\exp in\theta \\
\frac{J_{n}(z)}{z^{n}};\textrm{ }z &=&\sqrt{2}Er
\end{eqnarray*}%
If usually one associates $\exp in\theta $ to the homotopic n-loop the
corresponding Bessel function does not .Bessel functions scale as $%
J_{n}(z)\propto z^{n}$ when $z$ goes to zero and so could not describe a
loop winding around the puncture (origin) which is physically present as it
is trapped by the puncture .We therefore consider that positively oriented
loops ( $n\succ 0$) are described rather by the reduced Bessel function $%
\frac{J_{\zeta }(z)}{z^{n}}$ as $\lim_{z\rightarrow 0}\frac{J_{\zeta }(z)}{%
z^{n}}\neq 0.$ Negatively oriented loops ( $n\prec 0$) on the other hand are
descibed by $z^{n}J_{n}(z)$ .

Now that Bessel functions are supposed to realize homotopic loops in the z
variable .It is natural to look for the operator in the z variable which is
analogous to the winding number operator $W$

The operator that plays the role of the winding number operator; $W_{r}$ has
the form

\[
W_{r}=-\frac{1}{2}(z\frac{d}{dz}+(\frac{d}{zdz})^{-1}) 
\]

and it acts on the wavefunctions as follows

\begin{eqnarray*}
W_{r}\,\;\frac{J_{n}(z)}{z^{n}} &=&n\;\frac{J_{n}(z)}{z^{n}}\quad \\
\;W_{r}\,\;(z^{n}J_{n}(z)) &=&-n\;z^{n}J_{n}(z)
\end{eqnarray*}

The proof is as follows.

\begin{eqnarray*}
W\,_{r}\;\frac{J_{n}(z)}{z^{n}} &=&-\frac{1}{2}(z\frac{d}{dz}+(\frac{d}{zdz}%
)^{-1})\;\frac{J_{n}(z)}{z^{n}} \\
&=&-\frac{1}{2}z^{2}\frac{d}{zdz}\frac{J_{n}(z)}{z^{n}}+\frac{1}{2}\frac{%
J_{n-1}(z)}{z^{n-1}} \\
&=&\frac{1}{2}z^{2}\frac{J_{n+1}(z)}{z^{n+1}}+\frac{1}{2}\frac{J_{n-1}(z)}{%
z^{n-1}} \\
&=&\frac{1}{2}\frac{(J_{n+1}(z)+J_{n-1}(z))}{z^{n-1}} \\
&=&n\frac{J_{n}(z)}{z^{n}}
\end{eqnarray*}

Where in the fourth line we apply the recursion formula $%
J_{n+1}(z)+J_{n-1}(z)=2n$ $J_{n}(z)$ .There is another operator which
applies on reduced Bessel functions in a special way.It is

\begin{eqnarray*}
d_{m} &=&\frac{d}{zdz}\cdots \frac{d}{zdz}=(\frac{d}{zdz})^{m}\textrm{ ; }m\in
Z \\
d_{-\mid m\mid }d_{\mid m\mid } &=&d_{\mid m\mid }d_{-\mid m\mid }=1 \\
(-1)^{m}d_{m}\frac{J_{n}(z)}{z^{n}} &=&\frac{J_{n+m}(z)}{z^{n+m}}\textrm{ }%
;m\in N \\
d_{m}\textrm{ (}z^{n}J_{n}(z)) &=&z^{n-m}J_{n-m}(z);\textrm{ }m\in N
\end{eqnarray*}%
We thus have two operators acting on reduced Bessel functions $W_{r}$ and $%
d_{m}$ and these are the analogous ( $r$-components ) of $W_{\theta
}=-id_{\theta }$ and $\Pi _{\theta }(m)=\exp im\theta $

\section{ Exp$^{\textrm{s}}$ vs Bessel`$^{\textrm{s}}$:The correspondence Rules}

$\bigskip $

The important remark we make toward unraveling new properties of Bessel
functions (one in this paper ) is that ($\zeta =n$ The separation parameter)
is common a parameter to $\exp in\theta $ and to $\frac{J_{n}(z)}{z^{n}}$,
both describing the same object :The $n-$loop.,Hence we suggest and prove
that :Certain specific relationships ammong  exponentials will be
transported to reduced Bessel functions via the following correspondence
rules CR$^{`s\textrm{ }}$

$\bigskip \ $These CR$^{`s\textrm{ }}$are collected in the following tableau.(
The notation E$^{im\theta }$ means that $\exp in\theta $ acts as an operator
)

\begin{equation}
\begin{tabular}{|c|c|}
\hline
$\theta $-components & $r$ -components \\ \hline
Function & Function \\ \hline
$\exp in\theta $ $;-\pi \leq \theta \leq \pi $ & $\frac{J_{n}(z)}{z^{n}}$ \\ 
\hline
$1;(n=0)$ & $J_{0}(z)$ \\ \hline
Homotopic group elements & Homotopic group elements \\ \hline
$\Pi _{\theta }\left( m\right) =$E$^{im\theta }$ $;$ $m\in Z$ & $\Pi
_{r}(m)=(-1)^{m}(\frac{d}{zdz})^{m}$ $;m\in Z$ \\ \hline
$1;(m=0)$ & $1\equiv d_{-\mid m\mid }d_{\mid m\mid }=d_{\mid m\mid }d_{-\mid
m\mid }$ \\ \hline
Winding number & Winding number \\ \hline
$W_{\theta }=-i\frac{d}{d\theta }$ & $W_{r}=-\frac{1}{2}(z\frac{d}{dz}+(%
\frac{d}{zdz})^{-1})$ \\ \hline
\end{tabular}
\label{Tab1}
\end{equation}

Few remarks are in order at this stage .First note that the topological
invariant $z\partial _{z}$ gets a meaning in the new framework.That is we
have $r\partial _{r}=-2W_{r}+\Pi _{r}(-1)$ .It is a combination of the
winding number $W_{r}$ and the homotopy group element $\Pi _{r}(-1)$
.Second, note the global restriction on the angle $-\pi \leq \theta \leq \pi 
$.Relationships valid within truncated values of the angle or in another
interval other that the above will not map to corresponding relationships in
Bessel functions.To understand the globality of the restriction let us
consider the identity

\[
-i\theta =\sum_{m\in Z/0}(-1)^{m}\frac{\exp im\theta }{m}\textrm{ };-\pi \prec
\theta \prec \pi 
\]

This formula serves to define the interval of validity of the the angle $%
\theta $ as the right hand side only involve transportable quantities that
is $\exp im\theta $ and the accompanying constants $(-1)^{m};\frac{1}{m}$
which are invariant under transport .If on the other hand we shift the angle
to $\theta \rightarrow \theta +\pi $,we get a \textquotedblleft
bad\textquotedblright\ non tranportable formula

\[
-i\theta =\sum_{m\in Z/0}\frac{\exp im\theta }{m}-i\pi \textrm{ };0\prec
\theta \prec 2\pi 
\]

as the interval of definition is no longer $-\pi \prec \theta \prec \pi $

Finally we may note that the periodicity of the exponential is not
transported by CR$^{s}$as globality fixes the angle to be in a fixed
interval.

Applying the CR$^{s}$we get a series of interesting correspondences listed
in the following tableau. 
\[
\begin{tabular}{|c|c|}
\hline
$\theta $-components & $r$ -components \\ \hline
E$^{im\theta }\exp in\theta =\exp i(n+m)\theta $ & $(-1)^{m}(\frac{d}{zdz}%
)^{m}\frac{J_{n}(z)}{z^{n}}=$ \\ \hline
& $\frac{J_{n+m}(z)}{z^{n+m}}$ \\ \hline
E$^{im\theta }$E$^{in\theta }=$E$^{i(m+n)\theta }$ & $(-1)^{m}(\frac{d}{zdz}%
)^{m}(-1)^{n}(\frac{d}{zdz})^{n}=$ \\ \hline
& $(-1)^{m+n}(\frac{d}{zdz})^{m+n}$ \\ \hline
$1\exp in\theta =\exp in\theta $ & $(\frac{d^{2}}{dz^{2}}+(2n+1)\frac{d}{zdz}%
+1)\frac{J_{n}(z)}{z^{n}}=0$ \\ \hline
$\left[ -i\frac{d}{d\theta },E^{im\theta }\right] =mE^{im\theta }$ & $\left[
W_{z},(-1)^{m}(\frac{d}{zdz})^{m}\right] =$ \\ \hline
& $m(-1)^{m}(\frac{d}{zdz})^{m}$ \\ \hline
$(i\theta )^{n}$ ; $c$ number & $\ \frac{d^{n}}{d\lambda ^{n}}\frac{%
J_{\lambda }(z)}{z^{\lambda }}\mid _{\lambda =0}$ \\ \hline
$i\theta $ ; operator & $\sum_{m\in Z/0}\frac{(-1)^{m}}{m}(\frac{d}{zdz}%
)^{m} $ \\ \hline
$\exp i\lambda \theta $ & $\frac{J_{\lambda }(z)}{z^{\lambda }}$ \\ \hline
E$^{i\lambda \theta }$ & $(\frac{d}{zdz})^{\lambda }$ \\ \hline
$-i\frac{d}{d\theta }\exp i\lambda \theta =\lambda \exp i\lambda \theta $ & $%
z(J_{\lambda +1}+J_{\lambda -1})=2\lambda $ $J_{\lambda }(z)$ \\ \hline
$-i\theta =\sum_{m\in Z/0}(-1)^{m}\frac{\exp im\theta }{m}\textrm{ }$ & $%
\sum_{m\in Z/0}\frac{(-1)^{m}}{m}\frac{J_{m}(z)}{z^{m}}=-\frac{d}{d\lambda }%
\frac{J_{\lambda }(z)}{z^{\lambda }}\mid _{\lambda =0}$ \\ \hline
$;-\pi \prec \theta \prec \pi $ & $=J_{0}(z)\ln z-\frac{\pi }{2}N_{0}(z)$ \\ 
\hline
\multicolumn{2}{|c|}{The Unification Formula for Bessel Functions of
Different Orders} \\ \hline
$\exp i(n+\lambda )\theta =$ & $\frac{J_{n+\lambda }(z)}{z^{n+\lambda }}=$
\\ \hline
$\exp (-\lambda \sum_{m\in Z/0}(-1)^{m}\frac{\textrm{E}^{im\theta }}{m}\textrm{ }%
)\exp in\theta $ & $\exp (-\lambda \sum_{m\in Z/0}\frac{1}{m}(\frac{d}{zdz}%
)^{m})\frac{J_{n}(z)}{z^{n}}$ \\ \hline
etc & etc \\ \hline
$\vdots $ & $\vdots $ \\ \hline
\end{tabular}%
\]

Let us prove some correspondences to illustrate the CR$^{s}.$We take as
examples the one leading to the Bessel differential equation and that
leading to the recursion formula.The remaining correspondences present no
difficulties to be proved .Start from the trivial identity

\[
1\exp in\theta =\exp in\theta 
\]

where $1$ is the unit operator .CR`$^{s},$according to the tableau $\ref%
{Tab1}$ ,dictate the following expression in terms of Bessel functions%
\begin{eqnarray*}
1\exp in\theta &=&\exp in\theta \\
&\Downarrow & \\
\frac{d}{zdz}(\frac{d}{zdz})^{-1}\frac{J_{n}(z)}{z^{n}} &=&\frac{J_{n}(z)}{%
z^{n}} \\
&\Downarrow & \\
\left[ \frac{d}{zdz}(\frac{d}{zdz})^{-1}+\frac{d}{zdz}(z\frac{d}{dz})-\frac{d%
}{zdz}(z\frac{d}{dz})\right] \frac{J_{n}(z)}{z^{n}} &=&\frac{J_{n}(z)}{z^{n}}
\\
\left[ \frac{d}{zdz}(z\frac{d}{dz}+(\frac{d}{zdz})^{-1})-\frac{d^{2}}{dz^{2}}%
-\frac{1}{z}\frac{d}{zdz}-1\right] \frac{J_{n}(z)}{z^{n}} &=&0 \\
\left[ \frac{d^{2}}{dz^{2}}+\frac{1}{z}\frac{d}{zdz}+2\frac{d}{zdz}W_{z}+1%
\right] \frac{J_{n}(z)}{z^{n}} &=&0 \\
\left[ \frac{d^{2}}{dz^{2}}+(2n+1)\frac{d}{zdz}+1\right] \frac{J_{n}(z)}{%
z^{n}} &=&0
\end{eqnarray*}

In the second line we added and substracted the quantity $\frac{d}{zdz}(z%
\frac{d}{dz})$ and grouped terms such as to show the winding operator $W_{z}$
which acts on $\frac{J_{n}(z)}{z^{n}}$by singling out the index $n.$The
result is the Bessel differential equation .The proof of the second example
is as follows

\begin{eqnarray*}
-i\frac{d}{d\theta }\exp i\lambda \theta &=&\lambda \exp i\lambda \theta \\
&\Downarrow & \\
W_{z}\frac{J_{n}(z)}{z^{n}} &=&\lambda \frac{J_{n}(z)}{z^{n}} \\
-\frac{1}{2}(z\frac{\partial }{dz}+(\frac{\partial }{zdz})^{-1})\frac{%
J_{n}(z)}{z^{n}} &=&\lambda \frac{J_{n}(z)}{z^{n}} \\
z^{2}\frac{J_{\lambda +1}(z)}{z^{\lambda +1}}+\frac{J_{\lambda -1}(z)}{%
z^{\lambda -1}} &=&2\lambda \frac{J_{\lambda }(z)}{z^{\lambda }} \\
z(J_{\lambda +1}(z)+J_{\lambda -1}(z)) &=&2\lambda J_{\lambda }(z)
\end{eqnarray*}

The last formula is a well known recursion formula .

\section{\protect\smallskip A new property of Bessel functions}

The CR$^{`s}$gave us among various known recursions formulas or differential
equations a very specific and new formula

\begin{equation}
\frac{J_{n+\lambda }(z)}{z^{n+\lambda }}=\exp (-\lambda \sum_{m\in Z/(0)}%
\frac{1}{m}(\frac{d}{zdz})^{m})\frac{J_{n}(z)}{z^{n}}  \label{t3}
\end{equation}

This formula has been tested using different methods$.$It has been shown to
be true  by direct analytical computations in . \cite{Nadia} and by mapping
the integer order Bessel function differential equation to the real order
one in a paper yet to be submitted .On the other hand such mapping has been
successfully applied to Neumman`$^{s}$and Hankel functions as well with a
formula similar to the above and when applied to Polynomials such as Hermite
and Laguerre it gives deformed versions of these polynomials which are
generalization of the older one with certain shared properties .

\section{Outlooks}

In studying the topological action $b_{1}$ on the punctured plane we pull
out all the topological invariants realized in the $r$ as well as the $%
\theta $ variables .As a consequence this leads us to CR$^{`s}$ between
exponentials and Bessel functions$.$In this specific case CR$^{`s}$ could
have led us to discover the well known formula $\frac{J_{n}}{z^{n}}\propto
\int _{unit-circle}(2\tau )^{-n}\exp (\tau -\frac{z^{2}}{4\tau })$ $\frac{%
d\tau }{\tau }$ which summarize these rules ( the exponential in the
integrand has an essential singularity at the origin,hence the punctured
plane is singled out ,the map $unit-circle$$\rightarrow \tau ^{-n}$ maps the
circle to the $n$-loop and finally any deformation ( $homotopy$) of the
integration path is allowed as this is a property of the complex plane).The
above analysis is however more promising if applied to more general actions $%
b_{i}$ on manifolds of richer cohomology H$_{r}$(M)$\times $H$^{r}$(M) .In
this more general case new topological invariants will show up together with
the functions they act on and new CR$^{`s}$ between these new functions.

\end{document}